\documentstyle[editedbook,epsfig,psfig,epsf]{mq}

\def\cm2{cm$^2$ }
\def\se1{s$^{-1}$ }

\begin{opening}
\title{SS433: outflows here and there}
\author{Z. Paragi$^{1,2}$, I. Fejes$^2$, R.C. Vermeulen$^3$, R.T. Schilizzi$^{1,4}$, R.E. Spencer$^5$}
\author{A.M. Stirling$^6$}
\institute{$^1$ Joint Institute for VLBI in Europe, Postbus 2, 7990~AA Dwingeloo, The Netherlands \\
$^2$ F\"OMI Satellite Geodetic Observatory, P.O. Box 546, H-1373 Budapest, Hungary \\
$^3$ Astron, Postbus 2, 7990~AA Dwingeloo, The Netherlands \\
$^4$ Leiden Observatory, Post box 9513, 2300~RA Leiden, The Netherlands \\
$^5$ Jodrell Bank Observatory, University of Manchester, Jodrell Bank, Macclesfield, \\ Cheshire, SK11 9DL, United Kingdom \\
$^6$ CFA, University of Central Lancashire, Preston, PR1 2HE, United Kingdom \\}
\end{opening}

\runningtitle{4th Microquasar Workshop papers}
\runningauthor{Paragi et al.}

\begin{document}
\vspace{-0.5cm}
\begin{abstract}
{\small 
We summarize the results of VLBA and global VLBI observations of SS433 between
1995 and 2000. With these observations we resolve the inner jet of the source 
and identify an absorption region ($\sim$ 25 AU), the ``central radio gap".
The radio gap is caused by free-free absorption of the jet radio emission by a 
flattened outflow from the binary system. Radio emission is detected on 100 AU scales
perpendicular to the normal jets (``equatorial emission region"; ``equatorial outflow"). 
At some epochs the emission is smooth but compact features are frequently detected. 
We suggest that equatorial outflows may be common in microquasars.
}
\end{abstract}

\section{The radio inner jets}

The radio inner jets resemble those in AGN in that the jet length 
is roughly inversely proportional to the frequency, and a ``core-shift" is observed
between 1.6 and 5 GHz --- indicative of self-absorption \cite{Paragi_99}.
The differences are that the SS433 jet is ballistic,
it contains protons, and a large fraction of the jet material is
probably thermal. 

\begin{figure}[htb]
\centering
\psfig{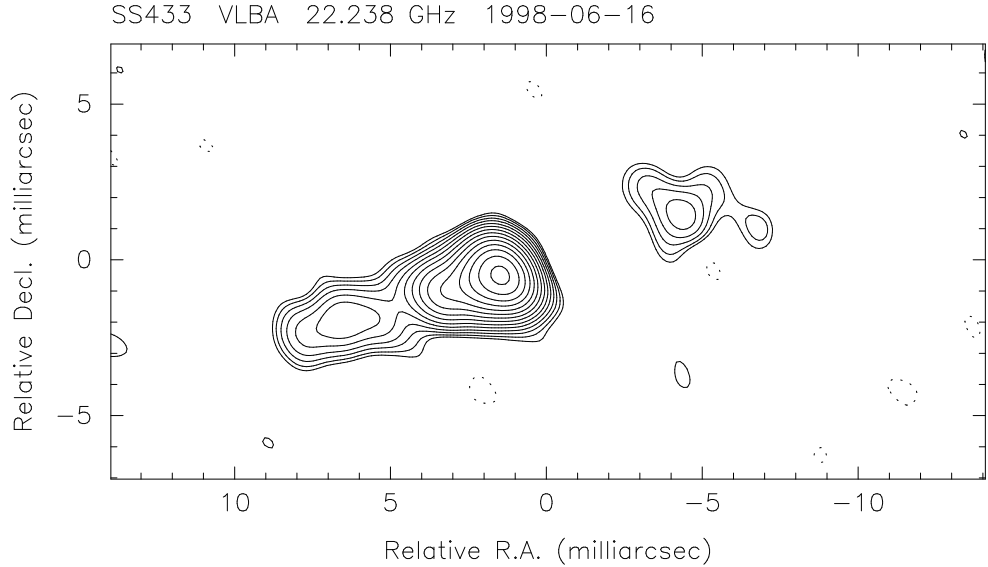}
\caption{VLBA image of SS433 at 22 GHz on 16 June 1998 (natural weighting).  
Contour levels increase by a factor of square root 2, the lowest contour is 
$\pm$1\% of the peak brightness (60.1 mJy/beam). The beam FWHM is 1.61$\times$1.15~mas,
its major axis is oriented at $33.3$ degrees.}
\label{fig:bp42d}
\end{figure}

\begin{figure}[htb]
\centering
\psfig{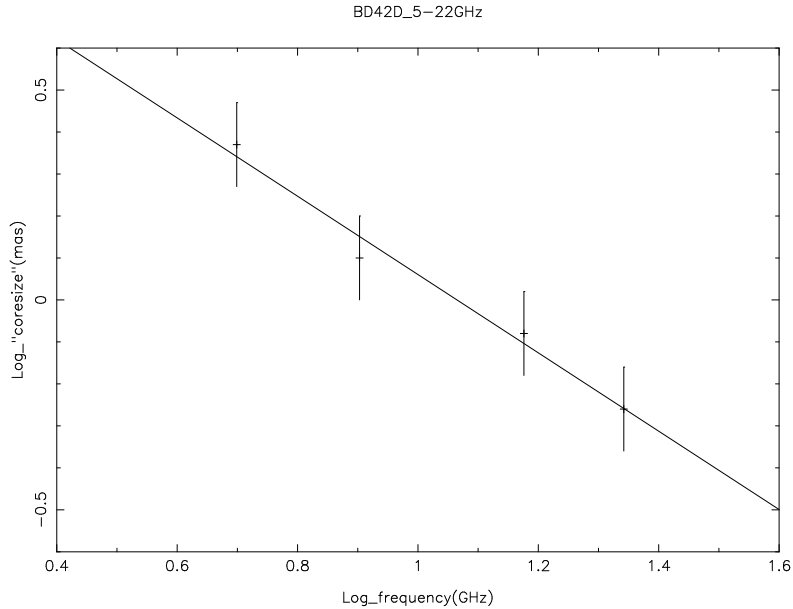}
\caption{Frequency dependence of the deconvolved Gaussian FWHM size of 
the main jet component on the approaching jet side on 16 June 1998, 
when SS433 was detected up to 22~GHz. The fitted ``core size" is 
proportional to $\nu^{-0.93\pm0.20}$. The brightness temperature remains 
roughly constant, it is order of $10^9$~K.}
\label{fig:logd-lognu}
\end{figure}

The approaching and receding jet sides are separated by a radio gap, as
was predicted by Stirling, Spencer \& Watson \cite{SSW97}. This
is caused by free-free absorption due to a disk-like outflow from the
central binary system. The brightness ratio of the two jet sides increases
with frequency, and at 22 GHz the receding jet is almost completely absorbed 
(Fig.~\ref{fig:bp42d}). The frequency dependence of the size of the 
mas-scale jet is
demonstrated in Fig.~\ref{fig:logd-lognu}. The Gaussian FWHM of the main 
jet feature on the approaching
jet side is proportional to $\nu^{-0.93}$ at an epoch when the inner jets
were detected up to 22 GHz. The apparent brightness temperature does not show a 
strong frequency dependence; it is $\sim$10$^9$~K on the approaching, and 
$\sim$10$^8$~K on the receding jet side. At 1.6 GHz the brightness ratio  
reflects only the Doppler-boosting effect and there seems to be no significant
free-free absorption on scales larger than $\sim$50~AU (assuming a distance 
of 5 kpc).

During large flares the inner jet region might disappear, and
pairs of plasmons are ejected from the system. There is indication 
for ongoing electron acceleration in these jet components (Paragi, Stirling
\& Fejes, these proceedings).

\begin{figure}[htb]
\centering
\psfig{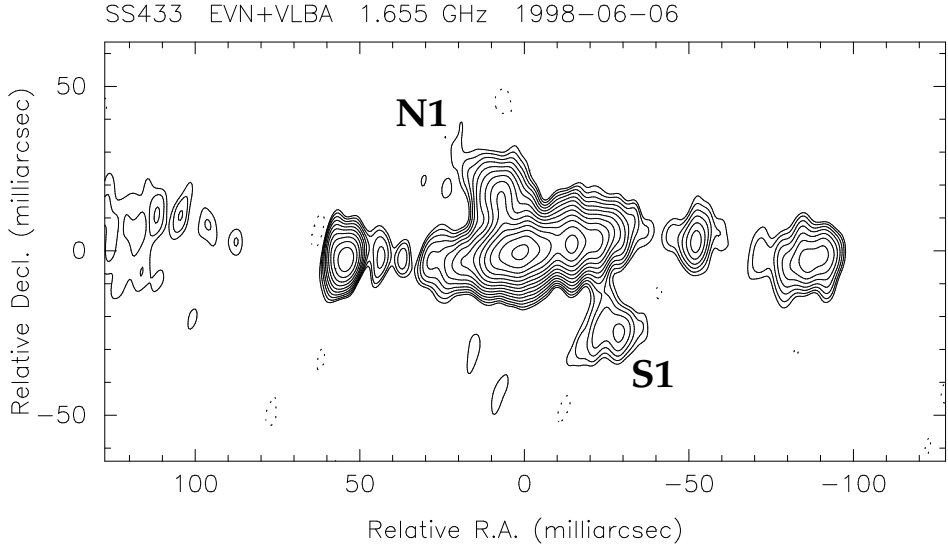}
\caption{Global VLBI image of SS433 at 1.6 GHz on 6 June 1998 (natural weighting). 
The Northern equatorial component (N1) was also detected at 5 GHz at two epochs
(Paragi, Fejes \& Szab\'o, these proceedeings). Its brightness temperature at 1.6 GHz
is $2\times10^{8}$~K. 
Contour levels increase by a factor of square root 2, the lowest contour is 
$\pm$1\% of the peak brightness (38.3 mJy/beam). The beam FWHM is 10.7$\times$3.76~mas,
its major axis is oriented at $-5.8$ degrees.}
\label{fig:gp017}
\end{figure}

\begin{figure}[htb]
\centering
\vskip 0.5cm
\psfig{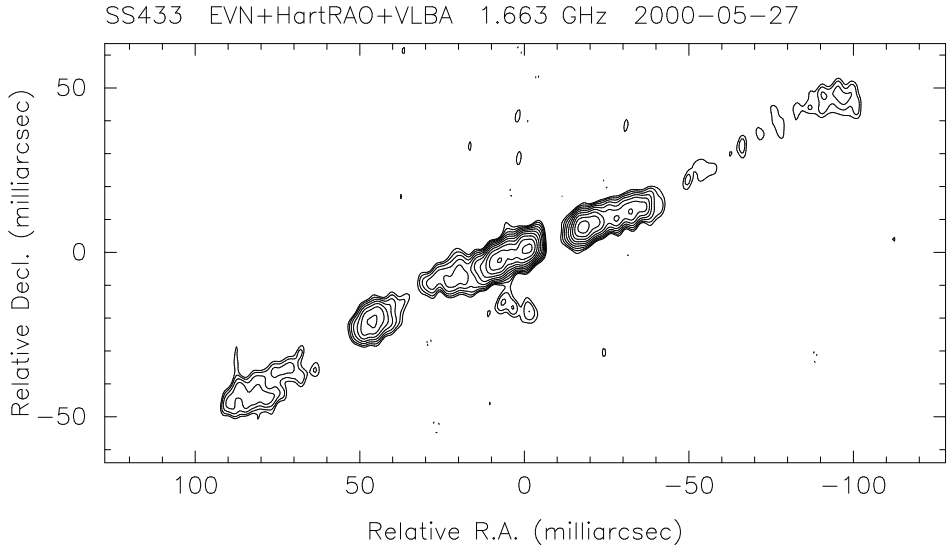}
\caption{Global VLBI image of SS433 at 1.6 GHz on 27 May 2000. We used uniform
weighting for this image, which results in the highest possible resolution
(and an increasement in the image noise at the same time). The inclusion of
Hartebeesthoek in the array improved the N--S resolution significantly. The 
equatorial component is separate from the jet, but appears very close to it. 
Contour levels increase by a factor of square root 2, the lowest contour is 
$\pm$4\% of the peak brightness (14.9 mJy/beam). The beam FWHM is 4.85$\times$2.82~mas,
its major axis is oriented at $-8.4$ degrees.}
\label{fig:gp025c}
\end{figure}

\section{The equatorial outflow}

We identified a radio emitting region perpendicular to the normal jets
\cite{Paragi_99} which was confirmed in a 1998 experiment
(Fig.~\ref{fig:gp017}).
As there were indications from observations, numerical simulations 
\cite{CM92} and theoretical works \cite{King00} that a significant 
fraction of the accreted matter 
leaves the system in its equatorial plane, we refer to this region as 
the Equatorial Outflow. We stress that the outflow is not 
constrained in a narrow plane, the term ``equatorial" refering to the fact 
that the region is not related to the radio beams of SS433 that 
emanate from the poles of the central engine. An independent group 
also observed this outflow (Blundell et al., these proceedings).

Emission from the equatorial outflow has been detected in seven VLBI 
experiments. The region is sometimes smooth, but compact, brighter features 
appear frequently. One of these components was observed to move away
from the centre at a projected speed of 1200$\pm$500~km/s (Paragi, Fejes
\& Szab\'o, these proceedings). This is significantly larger than the
estimated terminal wind speed of $\sim$300~km/s \cite{FBP97,Fab97}. 
Fabrika suggested that the faster disk 
wind ($\sim$1500~km/s) could develop shock waves into the slow 
($\sim$100~km/s) outflow from time to time, as the disk precesses. 
Another possibility is that the components are shocks resulting in a 
jet-ISM interaction. In Fig.~\ref{fig:gp025c} 
we see the emergence of a new equatorial
feature from the central region, close to the approaching Eastern jet.

Based on IR measurements Fuchs et al. (these proceedings) 
estimated a mass-loss rate
of $\sim$10$^{-4}$~M$_{\odot}$/yr assuming spherical symmetry. 
This result indicates that
most of the accreted matter leaves the system in this flow rather than
the radio jets. There are at least two other HMXBs where observations 
indicate a large-scale outflow embedding the binary system \cite{FHP99,Kanbach}.
For this reason we suggest that equatorial outflows may be common in 
HMXB microquasars.

\section*{Acknowledgments}
The European VLBI Network is a joint facility  of European and Chinese 
radio astronomy institutes funded by their national research councils.
The National Radio Astronomy Observatory is operated by Associated 
Universities, Inc. under a Cooperative Agreement with the National 
Science Foundation. We acknowledge partial financial support 
from the Hungarian Space Office (M\H{U}I), the Netherlands Organization 
for Scientific Research (NWO), and the Hungarian Scientific Research 
Fund (OTKA) (grant No. N31721 \& T031723). 
This research was supported by the European Commission's TMR Programme
``Access to Large-scale Facilities", under contract No.\ ERBFMGECT950012.
We acknowledge the support of the European Community - Access to Research
Infrastructure and Infrastructure Cooperation Networks (RADIONET, contract No.
HPRI-CT-1999-40003) action of the Improving Human Potential Programme.

\newpage

\phantom{\Large \bf Proper motion detected in the equatorial outflow of SS433 (poster)}
\phantom{\Large \bf Proper motion detected in the equatorial outflow of SS433 (poster)}
\hspace{-0.7cm}
{\Large \bf Proper motion detected in the equatorial outflow of SS433 (poster)}
\vspace{1cm}

\hspace{-0.7cm} 
{\large Z. Paragi$^{1,2}$, I. Fejes$^{2}$ \& A. Szab\'o$^{3}$} \\
{\small 
\hspace{-0.7cm}
$^{1}$ Joint Institute for VLBI in Europe, Postbus 2, 7990~AA Dwingeloo, The Netherlands \\
$^{2}$ F\"OMI Satellite Geodetic Observatory, P.O. Box 546, H-1373 Budapest, Hungary \\
$^{3}$ E\"otv\"os Lor\'and University, Department of Astronomy, P.O. Box 32, H-1518 Budapest, Hungary \\
}

\vskip 0.5cm

\begin{abstract}
{\small 
The equatorial outflow in SS433 has been observed in seven VLBI experiments
using the VLBA and the EVN. Emission is generally detected at only 1.6 GHz, 
but there are two epochs when bright features are observed also at 5 GHz.
One of these radio components is detected at three epochs, and appears to 
move outward from the central region at a projected velocity of 
$\sim1200\pm500$~km/s.
}
\end{abstract}

\setcounter{section}{0}
\setcounter{figure}{0}
\section{Introduction}

Since the Very Long Baseline Array observed SS433 in a multi-frequency
experiment in 1995, it has been known that microquasars may produce radio
emission not only in the well-studied jets, but also in their equatorial
region \cite{Paragi98,Paragi99}. 
Because the presence of an outflow had 
been suggested based on observations in the optical and in X-ray 
\cite{ZWI91,KOT96}, the radio emission was 
naturally attributed to this equatorial flow. Recent works in the 
UV 
\cite{DRG02} and IR regimes (Fuchs et al., these proceedings) also
support this scenario.
We initiated VLBA and global VLBI (EVN+Hartebeesthoek+VLBA) observations
to monitor changes in this region on milliarcsecond scales.

\section{Detected proper motion}

\begin{figure}[htb]
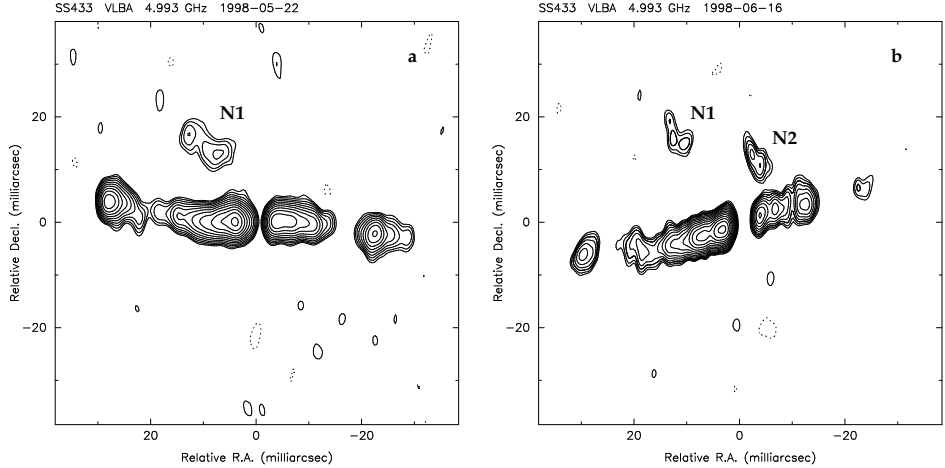

\centering
\vskip 1mm
\psfig{file=paragi2_1a.ps,width=6cm}
\hskip 3mm
\psfig{file=paragi2_1b.ps,width=6cm}
\caption{Component N1 as observed on 22 May 1998 ({\bf a}) and 
16 June 1998 ({\bf b}) at 5 GHz. Contour levels
increase by a factor of square root 2, the lowest contoures are $\pm$2\%
and $\pm$1\%. The peak brightness, beam FWHM size and orientation is
as follows: {\bf a)} 23.6 mJy/beam, 4.06$\times$1.86~mas, 0.6~degrees;
{\bf b)} 47.3 mJy/beam, 3.57$\times$1.51~mas, 0.1 degrees.
}
\label{fig:bp42cd}
\end{figure}

The existence of the equatorial emission region was confirmed by 1.6 GHz
global VLBI observations (see Fig.~3 of Paragi et al., these proceedings) 
on 6 June 1998. There were two radio components 
(indicated by N1 and S1), as in 1995, but their separation and position angle 
were different. N1 had a brightness temperature of $2\times10^{8}$~K, 
indicative of a non-thermal emission mechanism. It was also identified 
at 5~GHz on 22 May and 16 June 1998 (see Fig.~\ref{fig:bp42cd}).

Because the jets precess and there is no well-determined, compact feature
in the centre on milliarcsecond (mas) scales, it is difficult to find a good 
reference point in the source. We assumed that the binary is located 
in the middle of the radio gap \cite{Paragi99}.
The separation of N1
from this reference point increased significantly. By assigning a 1~mas error
to the position measurements (including the
uncertainty in the reference point), we determined an outward proper motion
of $\mu=0.14\pm0.06$~mas/day. If this is due to a bulk motion of matter,
this corresponds to an outflow velocity of (1200$\pm$500)/$\sin(i)$~km/s.

Changes in the region were further monitored in 2000 with a global VLBI
network (images not shown here). The results from these experiments can
be found in \cite{Paragi_EVN02}.

\section*{Acknowledgments}
The European VLBI Network is a joint facility  of European and Chinese 
radio astronomy institutes funded by their national research councils.
The National Radio Astronomy Observatory is operated by Associated 
Universities, 
Inc. under a Cooperative Agreement with the National Science Foundation.
We acknowledge partial financial support from the Hungarian Space
Office (M\H{U}I), the Netherlands Organization for Scientific Research (NWO), 
and the Hungarian Scientific Research Fund (OTKA) (grant No. N31721 
\& T031723). 
This research was supported by the European Commission's TMR Programme
``Access to Large-scale Facilities", under contract No.\ ERBFMGECT950012.
We acknowledge the support of the European Community - Access to Research
Infrastructure and Infrastructure Cooperation Networks (RADIONET, contract No.
HPRI-CT-1999-40003) action of the Improving Human Potential Programme.

\newpage

\phantom{\Large \bf Proper motion detected in the equatorial outflow of SS433 (poster)}
\phantom{\Large \bf Proper motion detected in the equatorial outflow of SS433 (poster)}
\hspace{-0.7cm}
{\Large \bf Electron acceleration in SS433 jet components (poster)}
\vspace{1cm}

\hspace{-0.7cm} 
{\large Z. Paragi$^{1,2}$, A. Stirling$^{3}$ \& I. Fejes$^{2}$} \\
{\small 
\hspace{-0.7cm}
$^{1}$ Joint Institute for VLBI in Europe, Postbus 2, 7990~AA Dwingeloo, The Netherlands \\
$^{2}$ F\"OMI Satellite Geodetic Observatory, P.O. Box 546, H-1373 Budapest, Hungary \\
$^{3}$ CFA, University of Central Lancashire, Preston, PR1 2HE, United Kingdom \\
}

\vskip 0.5cm

\begin{abstract}
{\small 
We present multi-frequency VLBA observations of SS433 during an outburst.
Our data suggests that electron acceleration is taking place
in the radio plasmons ejected from the central source. 
}
\end{abstract}

\setcounter{section}{0}
\setcounter{figure}{0}
\section{Structural changes and spectral properties}

\begin{figure}[htb]
\vskip 0.5cm
\centering
\psfig{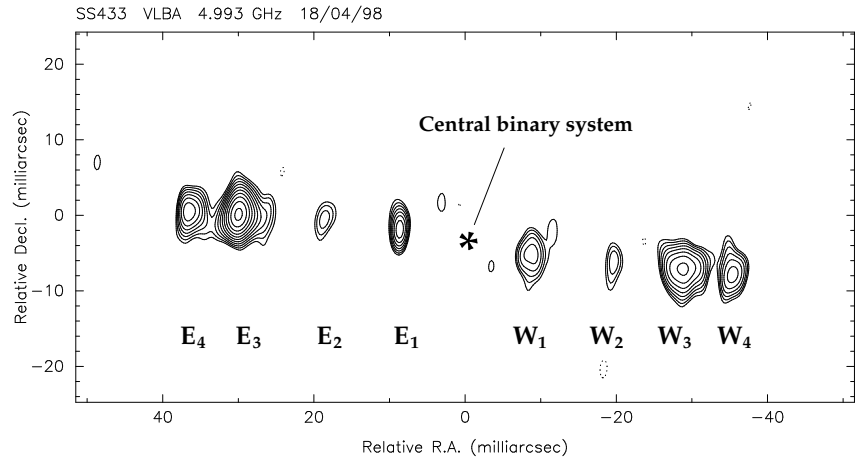}
\caption{Radio plasmons of SS433 observed with the VLBA on 18 April 1998
at 5 GHz. The brightest component (E3) had a brightness temperature of 
$2\times 10^{9}$~K.
Contour levels increase by a factor of square root 2, the lowest 
contour is $\pm$2.82\% of the peak brightness (109.7 mJy/beam). The beam 
FWHM is 3.71$\times$1.55~mas, its major axis is oriented at $-2.5$ degrees.
}
\label{fig:bp42b}
\end{figure}

\begin{figure}[htb]
\centering
\vskip 1mm
\psfig{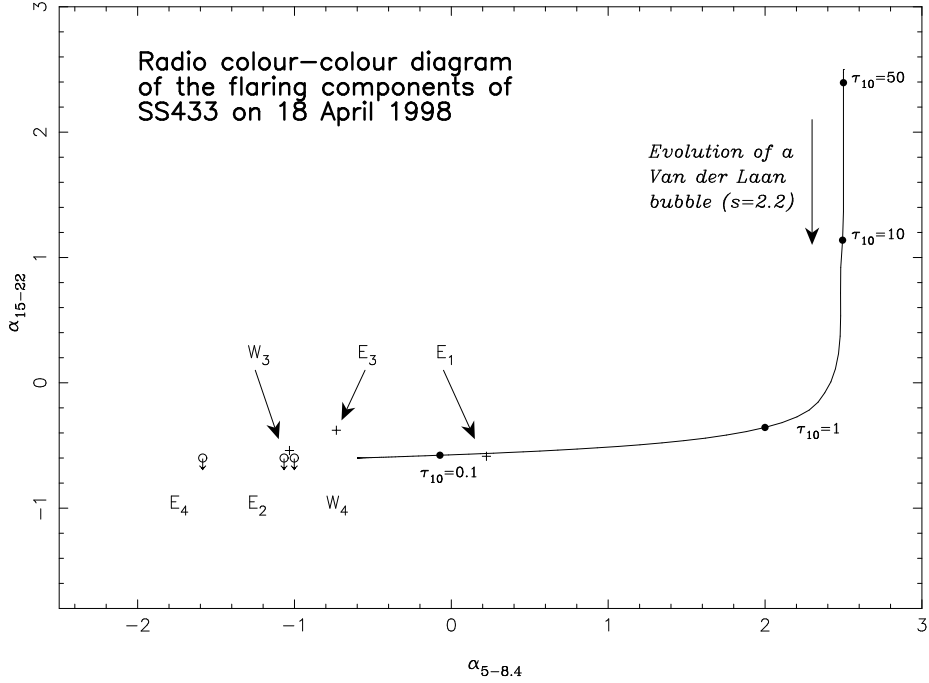}
\caption{The radio colour--colour diagram of SS433 jet components. The spectral 
evolution of a ``Van der Laan bubble" and its optical depth at the reference
frequency of 10 GHz is also shown. For some components there are only
upper limits available for the high-frequency spectral index. W2 was too faint 
even at 8~GHz; W1 lies outside the ranges shown. The electron energy spectral
index used in the model is $s$=2.2.
}
\label{fig:color}
\end{figure}

We observed SS433 with the VLBA (5, 8.4, 15 and 22 GHz) on 18 April 1998
during a large flare. The core-jet region of the source disappeared in
the flare. Instead we detected four pairs of plasmons ejected from the 
centre (Fig.~\ref{fig:bp42b}). Most of these components (except E1) had a 
very steep 
spectrum between 5 and 8.4 GHz. Interestingly, the spectrum between 
15 and 22 GHz was somewhat flatter. The spectral properties of the plasmons
are shown in Fig.~\ref{fig:color}, on the colour--colour ($\alpha_{15-22}-\alpha_{5-8.4})$ diagram \cite{col-col}.

The components are already in the optically thin phase of their evolution
at the observing frequencies. At this epoch the flare had not yet reached 
its maximum. The most straightforward explanation for an increasing
flux density during the optically thin phase is that a new generation 
of electrons was accelerated in the plasmons. A similar conclusion was
reached using single dish radio flux density measurements by Seaquist
et al. \cite{ERS82}.

\vskip -1mm

\section*{Acknowledgments}
The National Radio Astronomy Observatory is operated by Associated 
Universities, 
Inc. under a Cooperative Agreement with the National Science Foundation.
We acknowledge partial financial support 
from the Hungarian Space Office (M\H{U}I), Netherlands Organization for
Scientific Research (NWO), and the Hungarian Scientific Research Fund (OTKA) 
(grant No. N31721 \& T031723). 
We acknowledge the support of the European Community - Access to Research
Infrastructure and Infrastructure Cooperation Networks (RADIONET, contract No.
HPRI-CT-1999-40003) action of the Improving Human Potential Programme.

\end{document}